\documentclass[twocolumn,pra,superscriptaddress]{revtex4}
\pdfoutput=1
\usepackage{setspace}
\usepackage{amsfonts,amsmath,amssymb}
\usepackage{txfonts}
\usepackage{graphicx, color}
\usepackage{verbatim}
\usepackage{hyperref}
\usepackage{bm}

\begin{document}

\title{Reconciling the Classical-Field Method with the Beliaev Broken Symmetry Approach}

\author{Tod M. Wright} 
\affiliation{The University of Queensland, School of Mathematics and Physics, Brisbane, Queensland 4072, Australia} 
\author{Matthew J. Davis}
\affiliation{The University of Queensland, School of Mathematics and Physics, Brisbane, Queensland 4072, Australia} 
\author{Nick P. Proukakis}
\affiliation{School of Mathematics and Statistics, Newcastle University, Newcastle upon Tyne, NE1 7RU, UK}

\begin{abstract}
We present our views on the issues raised in the chapter by Griffin and Zaremba [A. Griffin and E. Zaremba, in \emph{Quantum Gases: Finite Temperature and Non-Equilibrium Dynamics}, N.~P. Proukakis, S.~A. Gardiner, M.~J. Davis, and M.~H. Szymanska, eds., Imperial College Press, London (in press)].  We review some of the strengths and limitations of the Bose symmetry-breaking assumption, and explain how such an approach precludes the description of many important phenomena in degenerate Bose gases.  We discuss the theoretical justification for the classical-field (c-field) methods, their relation to other non-perturbative methods for similar systems, and their utility in the description of beyond-mean-field physics.  Although it is true that present implementations of c-field methods cannot accurately describe certain collective oscillations of the partially condensed Bose gas, there is no fundamental reason why these methods cannot be extended to treat such scenarios.  By contrast, many regimes of non-equilibrium dynamics that can be described with c-field methods are beyond the reach of generalised mean-field kinetic approaches based on symmetry-breaking, such as the ZNG formalism.
\vspace{4\baselineskip}
\end{abstract}

\maketitle 

\section{Introduction}
\vspace{-1.0\baselineskip}
In Ref.~\cite{griffin_zaremba_chapter_12}, Griffin and Zaremba (GZ) have offered a critique of the so-called classical-field (c-field) method, comparing it to the Beliaev broken-symmetry approach to Bose superfluidity~\cite{beliaev_58,griffin_book_93}, and formalisms derived on this basis, such as the ZNG method~\cite{zaremba_nikuni_99,griffin_nikuni_book_09}.  A large part of their discussion concerns the validity of the separation of the Bose field operator into coherent (classical) and incoherent parts in practical applications of the classical-field approach, which they contrast with the separation into condensed and non-condensed parts in symmetry-breaking theories.  Regarding the application of these methods to the description of experimental systems, GZ focus on the collective modes of oscillation of finite-temperature BEC, which the ZNG formalism has reproduced with great success, and reason that the classical-field approach is unsuitable for the description of such phenomena.

Here we explain why the absence of a spontaneous symmetry-breaking assumption is not a failing of the classical-field approach, but is actually an important feature which ensures its generality and utility in describing many interesting scenarios of degenerate Bose-gas dynamics, including non-quasistatic growth of the condensate~\cite{weiler_neely_08}, regimes of matter-wave turbulence~\cite{berloff_svistunov_02}, and low-dimensional systems~\cite{proukakis_06b,bisset_davis_09}.  We also argue that the success of the ZNG method in modelling the collective oscillations of finite-temperature condensates does not appear to be due to many of the \emph{formal} consequences of the symmetry-breaking assumption cited in Ref.~\cite{griffin_zaremba_chapter_12}.  Although present implementations of the classical-field method lack the dynamical description of the above-cutoff atoms required to accurately reproduce some collective oscillations, there does not seem to be any fundamental reason why it cannot be extended to include these dynamics.  However, in our view the classical-field method is very useful even without such an extension, as it is the only existing methodology which can treat the non-equilibrium dynamics of the system in strongly fluctuating regimes.

\section{Symmetry Breaking \label{wright_symmetry_breaking}}
\vspace{-1\baselineskip}
A major point identified in Ref.~\cite{griffin_zaremba_chapter_12} is the absence of an \emph{a priori} distinction between condensed and non-condensed modes, such as that which accompanies symmetry breaking, in the classical-field approach.  In our view, this is not a failing of the method. As GZ note, the defining characteristic of Bose superfluidity is the emergence of a new thermodynamic variable on the superfluid side of the phase transition. The new variable that arises is, fundamentally, the superfluid velocity, which underpins Landau's famous two-fluid model of superfluidity~\cite{landau_lifshitz_book_87}.  This velocity field is in general attributed to the presence of an underlying condensate~\cite{pitaevskii_stringari_book_03}, and is determined by the \emph{gradient} of the condensate phase. A definite value for the phase [U(1) symmetry breaking] is thus not required to define the superfluid velocity; this was noted by Anderson~\cite{anderson_66}, who nevertheless advocated a description of superfluidity based on symmetry breaking, as it provides a unified description of both superfluidity and macroscopic interference phenomena in condensed systems~\cite{josephson_62}.  As is well known, true spontaneous symmetry breaking only occurs in infinite systems \cite{blaizot_ripka_book_86}, and for an isolated system of atoms, the particle-number superselection rule formally prohibits the appearance of a finite field expectation value $\langle \hat{\Psi}\rangle$~\cite{leggett_01}.  Nevertheless, field-theoretical calculations are typically formulated in a `restricted ensemble' in which a finite first moment of the field does exist~\cite{hohenberg_martin_65}; such a restricted ensemble can be introduced by adding symmetry-breaking terms to the Bose-field Hamiltonian~\cite{griffin_book_93}, which correspond to an external phase reference and thus lift the superselection rule (see Ref.~\cite{bartlett_rudolph_07} and references therein).  In this chapter we do not discuss the fundamental validity of this approach, but concern ourselves only with the pertinent \emph{operational} qualities of symmetry-breaking theories.  It is of course true that many profound and elegant consequences follow immediately from the assumption of Bose symmetry breaking~\cite{griffin_book_93}.  GZ note in particular that the symmetry-breaking approach makes it evident that the velocity field of the condensate exactly determines the local velocity of the total superfluid density --- which is in general not equal to the condensate density --- and yields a rigorous definition of this superfluid density~\cite{talbot_griffin_84}.  However, it does not appear to be {\em necessary} to appeal to symmetry breaking to define these quantities in general~\cite{leggett_book_06}.

Moreover, the natural relationship between the condensate and the superfluid density in the symmetry-breaking approach is of little consequence in the application of many formalisms based on symmetry breaking, such as the ZNG method~\cite{zaremba_nikuni_99,griffin_nikuni_book_09}.  GZ stress that the symmetry-breaking separation of the system into condensed and non-condensed (thermal cloud) parts in the ZNG method gives a `natural way of capturing the two-fluid nature of superfluids resulting from an underlying Bose condensate'.  However, in existing implementations of the ZNG method, the non-condensate atoms are treated in a single-particle (Hartree--Fock mean-field) description, and as such, the superfluid component is precisely the condensed component of the system (described by a generalised Gross-Pitaevskii equation).  One would expect that the implementation of a Bogoliubov-quasiparticle model of the thermal cloud (Chapter 7 of Ref.~\cite{griffin_nikuni_book_09}) would describe the small non-condensed contribution to the superfluid density in the comparatively straightforward case of a three-dimensional condensate.  However, scenarios of experimental interest in which there is a pronounced difference between the condensate and the superfluid density, such as the two-dimensional Bose gas --- in which thermal fluctuations erode long-range order but leave the superfluid density relatively unaffected~\cite{kosterlitz_thouless_73,popov_72,fisher_hohenberg_88} --- cannot be treated in the ZNG approach, precisely because such systems do not conform to the limiting case of a distinct condensate with well-defined quasiparticle excitations.

It is important to note that there are many such features of the condensed Bose gas which are fundamentally beyond a description in terms of a symmetry-breaking approach.  The symmetry-breaking ansatz assumes that the amplitude of the condensate orbital is a classical variable that, as GZ note, does not undergo any fluctuations, and cannot exhibit any higher-order correlations with the non-condensed component of the field.  By contrast, as one approaches the transition to the normal state the condensate population exhibits increasingly large fluctuations~\cite{campostrini_hasenbusch_01,davis_morgan_03,bezett_blakie_09a,wright_proukakis_11}, which require a description using \emph{non-perturbative} methods~\cite{capogrosso_giorgini_10}, and more generally particle-number conservation implies anticorrelation between the condensed and non-condensed populations~\cite{castin_chapter_01}.

GZ also review the consequences of the assumption of symmetry breaking for the excitations of the system.  In particular, they refer to the fact that the Bose broken symmetry leads to the condensate and non-condensate components sharing the same single-particle excitation spectra, and that the spectrum of density fluctuations of the system is also locked to the single-particle spectrum~\cite{griffin_book_93}.  These are indeed some of the most profound and elegant results of the symmetry-breaking approach.  However, approximate field theoretical formalisms based on symmetry breaking almost uniformly fail to reproduce these formal results.  This is the famous Hohenberg--Martin dilemma~\cite{hohenberg_martin_65, griffin_96}: all approximations for the self energies in the conventional Green's function approach either violate conservation laws (and thus do not yield an acoustic density-fluctuation spectrum), or violate the Hugenholtz--Pines theorem \cite{hugenholtz_pines_59,hohenberg_martin_65}, which requires a gapless single-particle spectrum.  As a result, the equivalence of the single-particle and density-fluctuation spectra is never realised in these approaches, and this is also true of the ZNG formalism.  By contrast, the so-called dielectric formalism~\cite{wong_gould_74}, an alternative perturbative approach which evades the Hohenberg--Martin dilemma and ensures coincidence of the two spectra, has not produced an accurate model of high-temperature collective oscillations~\cite{reidl_csordas_00}.  The undeniable success of the ZNG method in treating collective oscillations of finite-temperature condensates \cite{jackson_zaremba_01,jackson_zaremba_02c,jackson_zaremba_03,griffin_nikuni_book_09} therefore does not appear to be related to many of the {\em formal} features of the symmetry-breaking field-theoretic framework cited by GZ.

Finally, we note that although it does seem necessary to invoke symmetry breaking to provide a basis for a general field-theoretic approach to perturbation theory in the presence of a condensate~\cite{martin_dedominicis_64,hohenberg_martin_65}, the predictions of symmetry-breaking methods for the excitation spectrum, including the Bogoliubov, Hartree--Fock--Bogoliubov, and Beliaev--Popov approximations, are also obtained in explicitly number-conserving approaches (see references \cite{gardiner_97,castin_dum_98}, \cite{girardeau_arnowitt_59}, and \cite{morgan_00,andersen_04}, respectively).

\section{Basis of the Classical-Field Approximation}
The classical-field method is distinguished by the separation of the Bose field operator into two components: a part composed of single-particle modes which comprise the coherent region (or condensate band), and the remaining modes that constitute the complementary incoherent region (see Ref.~\cite{blakie_bradley_08}).  GZ contrast this distinction between the two regions with the division of the field operator into condensed (classical mean field) and non-condensed parts in the Beliaev method.  They note that the classical field $\Psi_{\mathbf C}(\mathbf{r})$ in the classical-field approach has (in contrast to the symmetry-breaking condensate of the Beliaev approach) `no special status' and that the classical-field separation is `artificial'.  It is indeed true that the separation of the Bose field into coherent and incoherent parts is not based on a strict physical distinction, and as such, $\Psi_{\mathbf C}(\mathbf{r})$ is not to be interpreted as being in any fundamental way physically distinct from the remainder of the Bose field as, for example, the condensed component of the field is in the Beliaev approach.  Nevertheless, this separation is an important component of the c-field method.

The division of the field in the c-field method can be motivated as follows.  The lowest-energy components of the interacting Bose field cannot be understood in terms of single-particle-like modes or excitations, as interactions strongly couple the single-particle modes of the corresponding non-interacting system (e.g., the harmonic-trap eigenstates).  In the limit of a well-defined condensate, this strong coupling is predominantly representative of the fact that the excitations of the condensed gas are Bogoliubov quasiparticles, rather than, e.g., dressed single-particle (Hartree--Fock) states~\cite{goldman_silvera_81}.  In more general situations, such a quasiparticle picture may be inapplicable, and the coupling of modes may be better understood in terms of density and phase fluctuations of an equilibrium quasicondensate (in low dimensional systems)~\cite{petrov_holzmann_00, petrov_shlyapnikov_00,andersen_alkhawaja_02,alkhawaja_andersen_02a,alkhawaja_andersen_02b}, or as a regime of strongly turbulent behaviour, such as is predicted to occur in the late stages of non-adiabatic condensation~\cite{kagan_svistunov_92a}.  Moreover, near the superfluid phase transition the field exhibits critical fluctuations~\cite{binney_book_92}, which herald the breakdown of perturbation theories and preclude the interpretation of the system in terms of a well-defined condensate and quasiparticle excitations in this regime~\cite{fedichev_shlyapnikov_98,morgan_00}.  In all of these scenarios, the low-energy portion of the field is characterised by correlations beyond a simple Hartree--Fock model of dressed single-particle states.

Fortunately, it is also the case, at least for moderately high-temperature regimes of degenerate Bose gases, that the single-particle modes that span the part of the field that exhibits non-trivial correlations are \emph{classically} occupied; i.e., they have mean occupations $\langle \hat{N}_k \rangle \gg 1$, such that their quantum fluctuations can reasonably be neglected.  As such, although the correlations and dynamics of this region may be quite non-trivial, the effects of \emph{quantisation} of the low-energy modes are comparatively subdued;  i.e., these correlations and dynamics result primarily from the multimode, self-interacting character of the field, not from its nature as a quantum field.

At higher energies, the occupations of field modes subside, so the highest-energy modes cannot be treated in a classical-field approach, and the excitations of the system also revert to a single-particle-like structure~\cite{gardiner_zoller_98}.  Provided that the return of the excitations to an essentially single-particle nature occurs at a \emph{lower} energy scale than that at which the classicality of mode occupations becomes violated (i.e., at which $\langle \hat{N}_k \rangle \lesssim 1$), one can introduce a division of the system into two parts (see Ref.~\cite{davis_wright_chapter_12}), such that one part contains all the non-trivial correlations and dynamics,  whereas the other contains all modes which have sub-classical occupations.  One thus identifies a component $\hat{\Psi}_{\mathbf{C}}(\mathbf{r})$ of the system which is potentially beyond any mean-field treatment, but which is, however, amenable to a classical-field description, while the complementary high-energy component $\hat{\Psi}_{\mathbf{I}}(\mathbf{r})$ can be treated in a simple mean-field approach~\cite{davis_blakie_06}.

The separation of the two components in the c-field approach is therefore indeed an artificial one, designed to allow for the application of the classical-field approximation to modes for which it is required, while not spuriously applying it to weakly-occupied modes for which it is not valid.  This discussion gives us guidelines as to the limits of the cutoff in the projected Gross-Pitaevskii equation (PGPE) / stochastic PGPE (SPGPE) approach: the cutoff should be at least of order $\mu\sim g |\phi|^2$ above the condensate eigenvalue, in order to include all quasiparticle structure and critical fluctuations in the classical field, while being not greater than $k_\mathrm{B}T$ above the condensate eigenvalue in order to ensure reasonably classical mode occupancies ($\langle \hat{N}_k \rangle \gtrsim 1$)~\cite{blakie_bradley_08} (see also Ref.~\cite{cockburn_thesis_10}).  Within these limits the choice of cutoff is arbitrary, and results should be practically independent of the precise choice of cutoff in this range~\cite{davis_blakie_06}.

It is important to note that there are two distinct, but closely related, classes of classical-field models. Microcanonical classical-field methods such as the PGPE involve closed-system, Hamiltonian equations of motion for the classical region of the Bose field, and neglect any coupling to the complementary incoherent component.  By contrast, the grand-canonical SPGPE~\cite{gardiner_davis_03} formalism --- which unifies the PGPE approach with the quantum kinetic theory of Refs.~\cite{gardiner_zoller_97,gardiner_zoller_98,gardiner_zoller_00} --- and the closely related Stoof stochastic Gross-Pitaevskii equation (SGPE)~\cite{stoof_99} involve explicitly stochastic equations for the classical field, with noise and damping terms which represent the effects of coupling to the above-cutoff region.  Although implementations of these methods to date assume the above-cutoff region to be in thermal equilibrium even when studying system {\em dynamics}~\cite{stoof_bijlsma_00,proukakis_03,weiler_neely_08,bradley_gardiner_08,cockburn_proukakis_09,cockburn_nistazakis_10,rooney_bradley_10,cockburn_nistazakis_11}, this is not a {\em fundamental} restriction of such formalisms~\cite{stoof_chapter_98,stoof_99,proukakis_jackson_08}.

\section{Equilibrium Correlations in the Classical-Field Approach}
As the classical-field approach is not based on a division of the field into condensed and non-condensed components, quantities such as the condensate must be inferred from the field correlations \emph{a posteriori}~\cite{blakie_davis_05a,proukakis_06b}.  In cases where a well-defined condensate exists, it can be found by applying the Penrose--Onsager definition of condensation~\cite{penrose_onsager_56} to the classical analogue of the one-body density matrix~\cite{blakie_davis_05a}.  More generally, properties of the system such as  quasicondensation and superfluidity can also be deduced from correlation functions of the field.

\subsection{Microcanonical Classical Fields}
The extraction of equal-time correlation functions from the PGPE is straightforward at equilibrium, as the nonlinear Hamiltonian evolution of the c-field samples the corresponding microcanonical density defined by the conserved energy and other first integrals of the system~\cite{blakie_bradley_08}.  Correlation functions of the field, such as the one-body density matrix, are thus formally given by averages over this microcanonical density, which can be approximated by averages over time.  It is important to note that, having extracted the Penrose--Onsager condensate orbital from a classical-field simulation, one can define a `fluctuation field', composed of the part of the field orthogonal to the condensate orbital, but taking into account the \emph{relative} phase between this field component and the condensate~\cite{wright_proukakis_11}.  This field is then the c-field analogue of the number-conserving non-condensate field operators introduced in symmetry-preserving quantum-field approaches~\cite{girardeau_arnowitt_59,gardiner_97,castin_dum_98,girardeau_98,morgan_00,gardiner_morgan_07}.  In this way, one can calculate [in addition to the coherent-region part of the non-condensate density $n'(\mathbf{r})$] the so-called \emph{anomalous} moments of the non-condensed component of the field~\cite{wright_proukakis_11} in the U(1)-symmetric microcanonical ensemble of the PGPE (see also Refs.~\cite{wright_blakie_10,cockburn_negretti_11,sinatra_witkowska_11}).  These correlations correspond to many-body processes, which yield corrections to the condensate-condensate and condensate--non-condensate scattering~\cite{proukakis_burnett_98,proukakis_morgan_98,hutchinson_burnett_00,morgan_00} and, in particular, to the chemical potential of the condensate.  An analysis of a three-dimensional condensate at finite-temperature equilibrium~\cite{wright_proukakis_11} shows that the anomalous averages can be significant, and moreover that the condensate obtained from a Penrose--Onsager analysis of the classical field appears to be consistent with the mean-field picture of the condensate as a nonlinear eigenfunction of a generalised Gross-Pitaevskii operator that involves pair and triplet anomalous moments~\cite{zaremba_nikuni_99, proukakis_burnett_98} (see Ref.~\cite{cockburn_negretti_11} for a related study of the pair anomalous average with the one-dimensional SGPE).
\vspace{0.25\baselineskip}

The extension of the ZNG method to describe the non-condensed component of the gas in terms of Bogoliubov quasiparticles (see Chapter~7 of Ref.~\cite{griffin_nikuni_book_09}) would --- as in the generalised mean-field kinetic treatments of Refs.~\cite{walser_williams_99,walser_cooper_01,kohler_burnett_02} --- include the pair anomalous average, and the attendant correction to the condensate's self interaction.  We note, however, that in present implementations of the ZNG method the anomalous average is neglected, despite not necessarily being small \emph{a priori}.  Moreover, it is not so clear how the higher-order correlations of the non-condensate fluctuations, such as the triplet correlations, could be built into the ZNG model.  The presence of these correlations in c-field equilibria underlines not only the naturally self-consistent nature of the Penrose--Onsager condensate in the classical-field equilibrium, but also the fact that the c-field method furnishes a \emph{non-perturbative} description of the field fluctuations.
 
\vspace{0.25\baselineskip}
We also note that as the PGPE is a fundamentally \emph{dynamic} equation of motion, it (as well as other Hamiltonian classical-field methods~\cite{brewczyk_gajda_07,sinatra_castin_07}) also gives access to (approximate) dynamic correlations at equilibrium~\cite{brewczyk_borowski_04,sinatra_castin_07,wright_ballagh_08,sinatra_castin_08,sinatra_castin_09,wright_bradley_09,wright_blakie_10,bezett_lundh_12,kulkarni_lamacraft_12}, as in the `Landau dynamics' approach to spin models~\cite{chen_landau_94}.  Importantly, this dynamical character is not merely introduced artificially to provide a means of sampling the thermodynamic ensemble (cf. Ref.~\cite{callaway_rahman_82}), but forms an approximation to the actual dynamics of the Bose field:  for large mode occupations, the evolution of the bosonic field is well described by a classical field equation, from which the quantum-field dynamics can be approximated by higher-order corrections in the inverse mode occupation~\cite{polkovnikov_03}.

\subsection{Grand-Canonical Classical Fields}
The stochastic GPEs~\cite{stoof_99,gardiner_davis_03} describe the c-field as a stochastic process, with correlations obtained in principle from averages over an ensemble of distinct trajectories.  In regimes where the underlying Hamiltonian part of the evolution behaves ergodically, one expects that a single trajectory will cover the appropriate thermodynamic ensemble densely, as in the usual Langevin equation approach to sampling thermal distributions~\cite{binney_book_92}.  One can therefore substitute time averages for ensemble averages at equilibrium.  However, in low-dimensional regimes, one must be mindful of the possible proximity of the underlying Hamiltonian part of the evolution to nearby integrable models (see, e.g., Ref.~\cite{cassidy_mason_09}) as, in general, the addition of noise and damping terms may not overcome the tendency of the system to remain `trapped' in quasi-regular regions of phase space~\cite{herbst_ablowitz_89}, preventing efficient exploration of the thermodynamic ensemble.  Correspondingly, studies of one-dimensional equilibrium systems with the Stoof SGPE explicitly consider an ensemble of distinct trajectories~\cite{proukakis_03,proukakis_06b,cockburn_proukakis_09,cockburn_gallucci_11,cockburn_negretti_11}.

\subsection{Fluctuations}
An important feature of the classical-field approach, noted by GZ, is its description of field fluctuations.  In general, it is only in the limit of a well-defined condensate that a clear division into a mean-field condensate and a complementary non-condensate part is valid.  As noted in Section~\ref{wright_symmetry_breaking}, the overall \emph{amplitude} of an otherwise well-defined condensate may undergo fluctuations, which become important as the system approaches the transition to the normal phase.  More generally, and particularly in low-dimensional systems, the concept of a condensate may not be useful in describing the system, and the appropriate `mean-field' approaches in such cases take the alternative route of describing the system in terms of density and phase fluctuations about an equilibrium \emph{quasicondensate}~\cite{petrov_holzmann_00,petrov_shlyapnikov_00,andersen_alkhawaja_02,alkhawaja_andersen_02a}.  As the classical-field method is \emph{not} based on the assumption of the existence of a well-defined condensate or quasicondensate, it is equally applicable to those regimes.  In particular, direct comparisons between the Stoof SGPE and the modified Popov theory of Andersen \emph{et al.}~\cite{andersen_alkhawaja_02,alkhawaja_andersen_02a} have shown good agreement between the two (for related studies see Refs.~\cite{simula_davis_08,bisset_davis_09,foster_blakie_10,cockburn_negretti_11,cockburn_gallucci_11,davis_blakie_12}).  In general, one expects that a true condensate, if present, is given by the part of the field which is both density- and phase-fluctuation suppressed~\cite{alkhawaja_andersen_02a,alkhawaja_andersen_02b}, and SGPE equilibria have been found to be consistent with this identification~\cite{cockburn_negretti_11,cockburn_proukakis_chapter_12}.

An important related question raised by GZ is to what extent a classical-field approach can provide an adequate description of the critical region near the phase transition.  It is well known that the transition to Bose condensation belongs to the universality class of the classical $XY$ model~\cite{campostrini_hasenbusch_01,kashurnikov_prokofev_01}.  As such, any classical $|\psi|^4$-type model, such as that considered in the PGPE, describes the universal characteristics of the Bose-condensation transition, and such classical models are commonly used to investigate the critical physics of the Bose gas \cite{kashurnikov_prokofev_01, prokofev_ruebenacker_01, arnold_moore_01}.  As such approaches treat the system without recourse to any assumption of spontaneous symmetry breaking, the issues of infrared divergences alluded to by GZ (see, e.g., Ref.~\cite{andersen_04}) do not arise (though such divergences are naturally cut off in the experimentally relevant case of harmonic trapping~\cite{posazhennikova_06}).  The new feature of the PGPE classical-field approach as applied to the harmonically trapped case, is that one expects, having properly taken account of the trapping potential, to also include \emph{non-universal} features peculiar to the system being studied.  Provided that the cutoff is chosen appropriately, the method should yield quantitative predictions for the system's critical behaviour, and indeed, the predictions of the PGPE approach for the $T_\mathrm{c}$ shift of the harmonically trapped, interacting Bose gas agree with experiment to within the present experimental uncertainty~\cite{davis_blakie_06}.

In light of this discussion, it is potentially confusing that the classical component of the Bose field operator in classical-field methods is sometimes described as an `order parameter' \cite{stoof_99,proukakis_jackson_08}.  Since by construction $\Psi_{\mathbf C}(\mathbf{r})$ describes a range of low-energy modes in addition to the condensate (should one be present), we agree that it is incorrect to think of $\Psi_{\mathbf C}(\mathbf{r})$ as a formal order parameter (except when the cutoff is low enough for this to coincide with the condensate \cite{duine_stoof_01}).  However, $\Psi_{\mathbf C}(\mathbf{r})$ appears as a field described by an effective time-dependent Ginzburg--Landau equation (particularly in the stochastic GPEs~\cite{stoof_99,gardiner_davis_03}), and it is unfortunately conventional in the relevant literature to refer to the field appearing in such descriptions as an order parameter~\cite{aranson_kramer_02}; this should of course not be confused with the order parameter associated with the Bose-condensation transition.

\subsection{Superfluidity}
GZ note that the symmetry-breaking approach relates the superfluid flow directly to the velocity field of the underlying condensate, and that this feature does not have to be inserted into the symmetry-breaking theory as an additional assumption.  We stress, however, that such an assumption is not required in implementing the classical-field method.  In general, the superfluid density is given by the response of the field to an applied velocity field (or phase twist), which leads to various correlation functions from which the superfluid density can be calculated~\cite{baym_chapter_69,fisher_barber_73,weichman_88,griffin_book_93,chaikin_lubensky_book_95}.  The superfluid density can be extracted from the c-field using these theoretical expressions~\cite{foster_blakie_10}, just as (for example) the superfluid density of the $XY$ model is calculated from classical Monte Carlo calculations~\cite{schultka_manousakis_94}.

It is true that each of the eigenmodes of the one-body density matrix calculated in the classical-field approach defines its own individual velocity field; but this would of course be true of any one-body density matrix of a Bose system, obtained in any approximation.  As GZ note, one expects that inasmuch as the system exhibits a well-defined condensate, it is this particular mode which defines the superfluid velocity field~\cite{leggett_book_06}.  From the velocity fields of all the eigenmodes of the one-body density matrix, one can construct a `hydrodynamic' velocity field, which corresponds to the total mass current~\cite{leggett_chapter_00}.  This hydrodynamic velocity field bears, of course, no \emph{a priori} relation to superfluidity.  However, in a system which contains a superfluid component which is significantly larger than the condensate --- i.e., one in which there is a significant contribution to the superfluid density from non-condensed atoms --- the associated supercurrent should constitute a contribution to the hydrodynamic current that matches the velocity field defined by the condensate orbital, and preliminary c-field simulations of quasi-two-dimensional superflows suggest that this is indeed the case~\cite{wright_davis_12}.

\section{Non-Equilibrium Dynamics in the Classical-Field Approach}
\subsection{Non-Equilibrium Correlations}
An important focus of Ref.~\cite{griffin_zaremba_chapter_12} is the viability of the classical-field method as a means of describing the non-equilibrium dynamics of Bose superfluids.  GZ correctly note that although the substitution of time averages for ensemble averages is reasonable for equilibrium systems, it is not applicable in non-equilibrium situations, in which one must in general explicitly consider averages over an ensemble of different classical-field trajectories.  These are distinguished by distinct choices of initial conditions for the field (and in the stochastic GPEs by distinct samples of the dynamical noise processes~\cite{stoof_bijlsma_01}), which may be sampled from an equilibrium thermal distribution~\cite{weiler_neely_08,bezett_blakie_09}, or from the Wigner distribution corresponding to the Bogoliubov vacuum in simulations starting from zero temperature~\cite{wright_ballagh_08}.  Although averaging over an ensemble of trajectories may in general lead to situations where there is no well-defined condensate, it does nevertheless seem at least reasonable to interpret individual classical-field trajectories as {\em representative} of individual experimental realisations of the system~\cite{duine_stoof_01,blakie_bradley_08,cockburn_nistazakis_10}. The individual trajectories, while having no direct formal meaning in, e.g., the truncated Wigner interpretation, certainly closely resemble the images seen in experiments.  Moreover, aside from constructing \emph{formal} correlation functions, the ensemble of trajectories does allow one to amass summary statistics for, e.g., decay rates of solitons~\cite{cockburn_nistazakis_10,cockburn_nistazakis_11} and vortices~\cite{rooney_bradley_10}, and frequencies of topological-defect occurrence~\cite{weiler_neely_08,damski_zurek_10}.  By contrast, mean-field approaches such as the ZNG method only describe a `mean' trajectory for the condensate in non-equilibrium evolution~\cite{jackson_proukakis_07,jackson_proukakis_09}, and give no information on the possible statistical dispersion of such quantities in experiments.

In \emph{quasi-equilibrium} dynamical regimes, one may find that the timescale on which the macroscopic dynamics of the condensate take place is sufficiently long, compared to the coherence timescale of thermal fluctuations, that one can unambiguously extract a non-equilibrium condensate and thermal cloud from a {\em single} classical-field trajectory by averaging over suitable intermediate timescales \cite{wright_bradley_09, wright_bradley_10}. However, no such obvious separation of timescales appears in general non-equilibrium scenarios, making it difficult, if not impossible, to identify a single, well-defined condensate mode during the evolution.  We emphasise that this is, in some sense, precisely the point: in general non-equilibrium situations one may not be justified in assuming \emph{a priori} that a well-defined condensate exists.

\subsection{Collective Oscillations}
A particularly important regime of non-equilibrium dynamics discussed by GZ is that of collective oscillations of finite-temperature condensates, the  accurate description of which has presented a profound challenge for theories of Bose-condensate dynamics since their first realisations in experiments~\cite{jin_ensher_96,jin_matthews_97}.  The results of extensive theoretical studies by many authors~\cite{hutchinson_dodd_98,bijlsma_stoof_99,alkhawaja_stoof_00,storey_olshanii_00,jackson_zaremba_02c,morgan_rusch_03,proukakis_jackson_08} reveal that the coupling between the oscillations of the condensate and those of the thermal cloud is of crucial importance for describing the anomalous frequency up-shift of the $m=0$ quadrupole oscillation at high temperatures.  A recent study of the collective oscillations in a classical-field (PGPE) approach~\cite{bezett_blakie_09} found increasing down-shifts of the $m=2$ oscillations with increasing temperature, consistent with previous studies \cite{hutchinson_dodd_98}, but did not observe the high-temperature up-shift of the $m=0$ mode (see also Ref.~\cite{karpiuk_brewczyk_10}).  As noted by the authors of reference~\cite{bezett_blakie_09}, and by GZ, this is perhaps not surprising, as the dynamics of the above-cutoff incoherent region were neglected in that study.

\section{Discussion}
We emphasise here that all dynamical processes \emph{within} the coherent region are described (in the classical-field limit) by the c-field methods.  It is only processes which couple {\em across} the artificial boundary to the above-cutoff incoherent region which are neglected in the PGPE approach.  In the stochastic-GPE approaches, these processes are described by the dissipative and stochastic terms that represent the coupling of the c-field to the above-cutoff region.  However, these methods do neglect the dynamic evolution of the above-cutoff region, and the influence of the c-field on these dynamics.  We stress that making a classical-field approximation for the low-energy modes of the gas does not in itself necessarily preclude constructing a model that includes these processes. The coupling between the two components is in principle provided for in Stoof's path-integral derivation of the SGPE~\cite{stoof_99,stoof_chapter_01}, and although the Gardiner SPGPE is derived on the basis of an explicit `tracing out' of the above-cutoff region~\cite{gardiner_davis_03}, its derivation could be extended to include a kinetic description of the above-cutoff atoms.  In such extensions, the above-cutoff region could be treated in a Hartree--Fock approach (as in current ZNG implementations), and its contribution to the non-condensate mean-field potential $2gn'(\mathbf{r})$ could easily be included in the c-field equation of motion.  (This contribution is typically neglected on the basis that it is flat over the coherent region at high temperatures~\cite{davis_blakie_06}, but would, as GZ note, be essential for the accuracy of the coupled dynamics.)  Such an approach should ultimately yield a method which includes both the `egalitarian' treatment of highly occupied modes in the classical region, while also describing the coupling of the superfluid (quasi-)condensate to the \emph{complete} remainder of the gas; i.e., it would include all the physics of the ZNG method, while additionally allowing for non-trivial fluctuations and correlations in the low-energy component.

However, GZ assert that the `natural' way to include the coupling between the condensate and the thermal cloud of non-condensed atoms is by means of a many-body perturbation-theory approach based on the symmetry-breaking concept.  Although it is clear that the ZNG method introduces the crucial features necessary to describe the particular non-equilibrium dynamics of collective oscillations, we do not agree that it is therefore preferable to pursue such a broken-symmetry approach to the non-equilibrium problem \emph{in general}.  It is conceivable that the ZNG method could form the basis for a generalised method in which fluctuations of, e.g., the phase of the condensate about the mean-field solution could be calculated from the fluctuation-dissipation theorem (as in Stoof's approach \cite{stoof_99}), and used to augment the ZNG predictions \emph{ex post facto}, thereby extending the ZNG approach to describe more general states of the Bose field in (quasi-)equilibrium regimes.  However, this approach would still be based on the \emph{a priori} assumption that an underlying well-defined condensate exists, which may not be justified in more general non-equilibrium scenarios.  Indeed, there are situations in which any approach based on symmetry breaking is incapable of yielding a description of the field dynamics which is even \emph{qualitatively} correct (see, e.g., Ref.~\cite{lobo_sinatra_04}).

In the view of GZ, the utility of the classical-field method is that it allows one to `address some non-trivial non-equilibrium problems without having to introduce a lot of formal machinery typical of field theoretic calculations based on the Beliaev formalism'.  We hope that this discussion has clarified why we do not regard the classical-field method as a mere means of simplifying field theoretical calculations, but as a distinct tool, with a broad range of applicability that is largely complementary to that of kinetic methods based on symmetry breaking, such as the ZNG formalism.    

\section*{Acknowledgements}
\vspace{-0.7\baselineskip}
\emph{The original version of this manuscript was prepared in response to a private critique of the classical-field approach to finite-temperature Bose gases, written by Allan Griffin.  The goal of these correspondences was to understand each other's viewpoints on the utility of classical-field techniques for Bose gases as compared to the Beliaev broken symmetry approach.  Allan wrote the first draft of his manuscript with the intention that it might appear, alongside a response, as a chapter of the edited volume \emph{Quantum Gases: Finite Temperature and Non-Equilibrium Dynamics}, N.~P. Proukakis, S.~A. Gardiner, M.~J. Davis, and M.~H. Szymanska, eds., Imperial College Press, London (in press).  Unfortunately Allan passed away before formally replying to the original response.  Eugene Zaremba kindly agreed to continue the discussion in Allan's absence.}

We would like to acknowledge Allan Griffin and Eugene Zaremba for many discussions over an extended period.  Tod Wright and Matthew Davis acknowledge funding from the Australian Research Council via the Discovery Projects program (DP1094025, DP110101047).  Nick Proukakis acknowledges funding from the UK EPSRC, the hospitality of the University of Queensland, and his visiting professorships at the University of Toronto and Queen's University.

\bibliographystyle{prsty}

\end{document}